# Quantifying nanoscale electromagnetic fields in near-field microscopy by Fourier demodulation analysis

*Fabian Mooshammer\*‡, Markus A. Huber‡, Fabian Sandner,*

*Markus Plankl, Martin Zizlsperger, and Rupert Huber\**

Department of Physics, University of Regensburg, 93040 Regensburg, Germany

\*E-Mail: fabian.mooshammer@ur.de, Fax: +49 941 943 4223
\*E-Mail: rupert.huber@ur.de, Fax: +49 941 943 4223

ABSTRACT:

**Confining light to sharp metal tips has become a versatile technique to study optical and electronic properties far below the diffraction limit. Particularly near-field microscopy in the mid-infrared spectral range has found a variety of applications in probing nanostructures and their dynamics. Yet, the ongoing quest for ultimately high spatial resolution down to the single-nanometer regime and quantitative three-dimensional nano-tomography depends vitally on a precise knowledge of the spatial distribution of the near fields emerging from the probe. Here, we perform finite element simulations of a tip with realistic geometry oscillating above a dielectric sample. By introducing a novel Fourier demodulation analysis of the electric field at each point in space, we reliably quantify the distribution of the near fields above and within the sample. Besides inferring the lateral field extension, which can be smaller than the tip radius of curvature, we also quantify the probing volume within the sample. Finally, we visualize the scattering process into the far field at a given demodulation order, for the first time, and shed light onto the nanoscale distribution of the near fields and its evolution as the tip-sample distance is varied. Our work represents a crucial step in understanding and tailoring the spatial distribution of evanescent fields in optical nanoscopy.**

Focusing electromagnetic waves at a sharp scanning probe tip lies at the heart of a variety of successful nanoscopy techniques[1–9]. Thereby, the propagating far fields are converted into evanescent near fields around the tip apex inducing a strong localization of the radiation on subwavelength scales accompanied by a strong field enhancement. In scattering-type scanning near-field optical microscopy (s-SNOM), the light-matter interaction at the nanoscale is probed by collecting the far-field radiation scattesred out of the tip-sample junction. These fields can be analyzed, and the nanoscale dielectric properties of the material can be inferred by accounting for the non-trivial response function of the scattering process.

Using state-of-the-art analytical or numerical models[10–15] to describe the scattering of light from the tip-sample system, one can predict the spectral field amplitudes and phases for a given dielectric function of the sample. Over the last decades, these models have proven extremely successful in the mid infrared[16] – the natural frequency window for low-energy excitations in condensed matter and vibrational bands of molecules. Prominent near-field applications in this spectral regime include studies of nanostructures[17–20], surface polaritons[21–25], phase transitions[26–29], and chemical composition[30,31]. Recently, the capabilities of near-field microscopy have been extended to the retrieval of dielectric functions of layered structures[32,33] and liquids[34].

For a quantitative understanding of mid-infrared nanoscopy it is important to determine the involved nanoscale evanescent fields. Hence, a microscopic electrodynamical treatment of the propagating and confined electric fields of the complete tip-sample system – ranging from the near- to the far-field region – is needed. Numerical approaches, such as the finite element or finite difference methods[35–42], can, in principle, fulfill these requirements for a static position of the tip above the sample. Yet, in near-field microscopy the tip typically oscillates periodically above the sample, leading to a modulation of the fields at this tapping frequency and its harmonics. It has been commonly accepted that smaller volumes within the sample can be probed if the scattering signals are evaluated at higher harmonic demodulation orders[18,32,33]. Pushing the limits of the lateral and out-of-plane resolution of SNOM and nano-tomography, however, requires a quantitative knowledge of the exact spatial distribution of the demodulated evanescent fields and their dependence on the tapping amplitude and the demodulation order – a key challenge that has so far remained elusive.

Here, we resolve this fundamental challenge by mapping the demodulated near fields within and outside of the sample for different dielectric functions, tapping amplitudes and tip radii. Furthermore, we directly correlate the decay characteristics of the near fields with the scattered far fields. The simulation procedure is illustrated in Figure 1. We numerically solve Maxwell's equations using the finite element method for a geometry as shown in Figure 1a. A conical gold tip (length, $l = 20$ µm) with a spherical apex (radius, $r = 25$ nm) is placed above a dielectric medium, which is described by the dielectric function of silicon (see Methods for further simulation details). Tip and sample are illuminated with a monochromatic plane wave of mid-infrared light (frequency, $\nu = 30$ THz; wavelength, $\lambda = 10$ µm) at an incidence angle of 30° with respect to the sample surface in order to mimic the experimental geometries of state-of-the-art near-field experiments. A representative distribution of the resulting out-of-plane electric field component $E_z$ is given in Figure 1b. This field distribution, however, corresponds to a static position of the tip. To account for the tapping motion (see inset in Figure 1a), we repeat the simulation for various heights $h$ of the tip above the sample.

Figure 1c depicts the distributions of $|E_z|$ close to the tip apex for five representative heights. The field enhancement peaks directly beneath the apex – a feature that becomes even more pronounced when the tip is in close proximity to the sample surface. To model a full oscillation cycle of the tip, we assume $h$ to be a sinusoidal function of time $t$ (Figure 1d):

$h(t) = A \cdot \left(1 + \cos(2\pi f_{\text{tip}} t)\right) + d$. Here, $A$, $f_{\text{tip}}$, and $d$ represent the tapping amplitude, the tapping frequency, and the minimal tip-sample separation, respectively. We discretize $h(t)$ in intervals of 1 nm to simulate the electric field distribution for each time step of the tip oscillation.

A representative temporal evolution of the electric field at the position directly below the tip apex at the sample surface (Figure 1e) highlights the strong field "bursts" for small values of $h$. Consequently, a sinusoidal motion of the tip modulates the electric field not only at the tip tapping frequency $f_{\text{tip}}$, but also at its harmonics of order n (see Figure 1f). We label these emerging spectral components of the demodulated electric field at a specific point in space as $\tilde{E}_n$, where $\tilde{E}_n = |\tilde{E}_n| e^{i\varphi_n}$ contains the amplitude $|\tilde{E}_n|$ and the phase $\varphi_n$. In typical near-field experiments, the scattered radiation is collected with a parabolic mirror and the total electric field or intensity at the detector is demodulated at different orders n as the scattering signal $s_n$. These far-field amplitudes are then utilized to retrieve information about the nanoscale near-field interaction of the tip-sample system. For a given lateral tip position, only a single value for $s_n$ is obtained corresponding to an average response of the volume that is permeated by the near fields. In other words, it is generally assumed that the detected far-field quantities $s_n$ are linked to the average light-matter interaction in the vicinity of the tip apex. With our Fourier analysis, in contrast, we do not only access $s_n$, but also the demodulated fields $\tilde{E}_n$ on the nanoscale. Thereby, we can quantify the extension of the demodulated near fields, which are responsible for the far-field responses $s_n$, in all spatial dimensions, for the first time. Furthermore, we successfully visualize the light scattering from the tip to the detector and directly correlate the near fields with the measured far fields.

In the first step, we focus on the demodulated fields $\tilde{E}_n$ in the vicinity of the tip apex. Figure 2a depicts the result for the four lowest demodulation orders, which are obtained by the procedure outlined in Figure 1 (see Methods for further details). Note that the novel Fourier demodulation analysis accounts for the complete tapping motion of the tip. Consistently, the upper and lower points of inflection of the tip motion (white outlines) govern the field pattern. The field amplitudes $|\tilde{E}_n|$ are strongly localized about the volume that the tip apex covers during one oscillation cycle. The highest amplitudes concentrate at the lower point of inflection. Additionally, the pattern becomes more complex for increasing demodulation order n resulting in n nodes and n+1 "lobes" along the out-of-plane direction. Most importantly, the foremost lobe interacting with the sample becomes more tightly localized for increasing demodulation order. This behavior can be quantified using the field profiles in Figure 2b, which were extracted along the dashed line in Figure 2a. For increasing order n, the lateral full width at half maximum (FWHM) $2\Gamma$ of the field profiles decreases continuously. The exact values of $\Gamma$ – a direct measure for the achievable lateral resolution – as a function of the tip tapping amplitude $A$ and demodulation order n are depicted in Figure 2c. The extension of the fields can be tuned over a wide range of several tens of nanometers. Remarkably, these fields can be confined to length scales even smaller than the tip radius (see shaded region in Figure 2c and Supporting Information). These results corroborate that a spatial resolution of near-field microscopy on the single-nanometer level may become possible with extremely sharp, custom-tailored tips[43]. Approximating the tip geometry by a set of monopoles within the framework of the finite-dipole model[11] leads to qualitatively similar demodulated field distributions $\tilde{E}_n$ (see Supporting Information). Yet, the lateral confinement of the near fields is not reproduced quantitatively.

Our novel Fourier demodulation analysis of the near fields also allows us to quantify the penetration profile into the sample (Figure 2a, below the white solid line) – a crucial prerequisite for quantitative nano-tomography. Phenomenologically, the variation of the probing volume for different demodulation orders of the far-field response has been successfully used to incorporate tomographic sensitivity in s-SNOM[18,32,33]. However, a precise understanding of the relevant field distributions within the sample has been missing so far. Therefore, we study the region close to the sample surface (highlighted in Figure 2a) in more detail (Figure 3). This allows us to pinpoint the extension of the demodulated fields $\tilde{E}_n$ within the sample, for the first time. We find that the field amplitudes are strongly localized at the surface and distributed almost hemispherically around the lateral position of the tip apex (see Figure 3a). Note that for finite tip-sample offsets $d$, the distribution of the demodulated fields within the sample takes a qualitatively different shape (see Supporting Information). For increasing order n, the confinement of the fields to the surface becomes stronger (indicated by the black 1/e decay contour lines) – a fact that has been known empirically[18,32,33] from the far-field amplitudes $s_n$.

The evanescent character of the nanoscale field amplitudes $|\tilde{E}_n|$ can be investigated in a more quantitative fashion by extracting decay profiles along the out-of-plane direction (see dashed line in Figure 3a). Remarkably, the decay is not purely exponential, but features a second, slower component, which follows a power-law scaling (see Figure 3b). The fast decay takes place within the first few nanometers and the slower decay persists for tens of nanometers. We quantify both scales by the characteristic lengths at which the normalized fields have decayed to a value of 1/e or 0.1, respectively (see inset in Figure 3b). A summary of both quantities in Figure 3c reveals that the fast decay is only weakly affected by the experimentally accessible turning knobs such as the tapping amplitude or the demodulation order. In contrast, the slower decay length can be varied over probing depths of several nanometers. This tunability of the probing volume within the sample forms the basis for nano-tomography[18,32,33], which is essential for investigating the dielectric properties of layered samples or buried objects[44–47]. In the Supporting Information, we also corroborate that the field decay depends strongly on the tip radius $r$.

Our discussion has so far been centered around the evanescent fields $\tilde{E}_n$ below the tip as extracted with our novel Fourier demodulation analysis. Next, we focus on the scattering process of these nanoscale near fields from the tip apex into the macroscopic far field. This allows us to directly correlate our results with the experimentally accessible, demodulated signal $s_n$. Therefore, we plot the distribution of the field components $\tilde{E}_n$ within the full simulation volume (see Figure 4a). We observe an emission pattern that is reminiscent of the one created by a dipole within the tip-sample junction, exhibiting a periodicity corresponding to the wavelength λ of the incident light. Note that the pattern of the field component $\tilde{E}_1$ is slightly asymmetric and that there are a few residual sources of emission at the side and top parts of the metallic tip, which lead to an additional far-field background. By increasing the demodulation order (see $\tilde{E}_2$ in the right panel of Figure 4a), the emission pattern becomes perfectly symmetric and point-like around the nanojunction formed by the tip apex and sample. We note that the noise in these images is due to the numerical accuracy limits of the simulation and increases for higher demodulation orders since the process requires the extraction of decreasing differences of the field amplitudes.

With these insights, we can relate the demodulated electric near fields $\tilde{E}_n$ to a quantity that is readily measurable in the far field – the intensity on a detector (see Figure 4b) from which the scattered amplitudes $s_n$ can be inferred. Experimentally, the relation of near and far fields has often been probed in so-called retraction/approach curves, where the minimal tip-sample distance $d$ during the tip oscillation is increased/decreased and the characteristic decay of $s_n$ is measured. In Figure 4c, we simulate this scenario. The circles represent the near-field amplitudes $\langle|\tilde{E}_n|\rangle_V$ averaged over the probing volume $V$ (see "probing volume" in Figure 4b) as determined from data similar to Figure 3 (for representative field distributions with finite tip-sample distance $d$, see Supporting Information). We now directly compare these data to the far-field amplitudes $s_n$ (Figure 4c, red line). To this end, different quantities, such as the power passing through a dome surrounding the tip apex in air (see "far field" in Figure 4b), are evaluated and an averaged far-field retraction curve is obtained, represented by the red line. We verified that a variation of the geometry of the dome and an alternative evaluation of the surface charge density[38] induced into the tip as a measure for the scattered field yield identical results as indicated by the error margins (for further details see Supporting Information). The perfect agreement underlines that the spatially resolved near-field amplitudes $|\tilde{E}_n|$ obtained by our novel technique indeed govern the measured far-field response $s_n$.

Experimental retraction curves are typically used as a measure of the lateral resolution or the decay of the fields within the sample. Comparing Figure 4c and Figure 3b, however, it is apparent that – even though the mean near-field amplitudes $\langle|\tilde{E}_n|\rangle_V$ can be directly related to the far-field amplitudes $s_n$ – the decay profile of $|\tilde{E}_n|$ within the sample and the one obtained in a retraction curve are clearly different. In the following, we will show that retraction curves are actually a convolution of several effects. We first take a step back and examine the electric field $E_z$ at a fixed tip height $h$ (see inset in Figure 4d). This allows us to evaluate the field directly underneath the tip apex and at the surface by extracting line profiles (see inset in Figure 4e), from which we can quantify their width and magnitude (see Figure 4e). Tracing the width $2\Gamma$ of the field profiles (see Figure 4d), we identify two regimes: For tip-sample separations larger than the tip radius ($d > r$), the field below the apex is hardly affected by the sample and its width remains constant (green curve). In this regime, the field profile at the surface grows linearly in diameter with distance $d$ (blue curve) and its magnitude decays exponentially (see dashed lines in Figure 4e). In contrast, for $d < r$, there is a square-root-like increase of the width, which is in agreement with previous reports[48,49]. This strong localization of the fields is responsible for the drastic increase of their magnitude.

For the demodulated fields $\tilde{E}_n$, the behavior is very similar (see Figures 4d,e), but additionally they become more strongly localized and feature a faster decay with increasing order n. Furthermore, the decay profiles deviate strongly from the predictions of the finite-dipole model as an approximation of the field distribution close to the tip apex (see Supporting Information). At the surface, the width of the distribution of $E_z$ extracted for a fixed offset $d$ represents a lower boundary for the FWHM of $|\tilde{E}_n|$. Roughly speaking, the latter represents a field average over a range of tip heights $h \geq d$, owing to the additional tapping motion of the tip. A retraction curve is then the result of an interplay of probing a continuously increasing volume and the rapidly decreasing field strength therein. The exact decay length measured in experiments is also strongly affected by the geometry of the tip, especially its radius $r$.

Despite the complex evolution of the fields beneath the apex, we verify that for a given tip geometry only the optical response of a sample governs the decay length of a retraction curve. In our analysis, we require a positive real part of the dielectric function ($\varepsilon_1 > 0$) and a vanishing imaginary part of the dielectric function ($\varepsilon_2 = 0$) of the samples because strong polariton modes, for example, can affect approach curves in a non-trivial way[50]. By comparing two dielectrics – silicon and diamond – we find a clear correlation between their respective retraction curves. To this end, we determine the characteristic 1/e decay lengths at a given demodulation order and tapping amplitude for both materials. By plotting such pairs of decay lengths associated with each of the two materials for a series of demodulation orders and tapping amplitudes, we obtain a clear linear dependence (see Figure 4f). In the Supporting Information, we corroborate this direct proportionality of the decay lengths by experimental approach curves recorded on four different materials. The scaling factor between the decay lengths of different materials should be governed by their dielectric functions. A determination of the exact functional dependence will be pursued in future studies, where additional experimental and numerical data are required.

Recent alternative advances in mid-infrared nanoscopy have also reconstructed the near-field interaction as a function of the tip-sample distance. This has been achieved by either utilizing a Fourier series of the scattered amplitudes recorded at various demodulation orders[51,52] or without relying on any demodulation at all, instead exploiting the so-called peak force tapping mode[53,54]. We expect that the effective near-field interaction extracted from the scattered fields can be understood analogously to the discussion of the electric fields $E_z$ in Figures 4d,e. The interaction obtained at a given tip-sample distance $d$ should then be directly linked to the spatially averaged near fields within the sample. Combining this experimental approach with the theoretical framework presented here, quantitative nano-tomography of layered samples could also be within reach.

In conclusion, the novel Fourier demodulation analysis provides us with the full electric field distribution at an oscillating metallic tip of a near-field microscope in all spatial dimensions – ranging over various length scales from the vicinity of the apex to the far field. The new insights into the properties of the evanescent fields serve as a crucial prerequisite for performing quantitative nano-tomography and for expanding the limits of the lateral resolution in near-field microscopy. Tailor-made tips may also allow for polarization-shaped near fields[55] with nanometer precision – even below the tip radius of curvature. In the future, our approach can be extended to probe the nanoscale dielectric properties of layered sample structures and even atomically thin materials[22,23] as well as ultrafast near-field microscopy with single- and few-cycle light pulses[3–6,18,24,56–59]. Finally, a combination of our Fourier demodulation analysis with fully quantum mechanical theories[60] could pave the way to a consistent description of light scattering from atomically sharp tips[8].

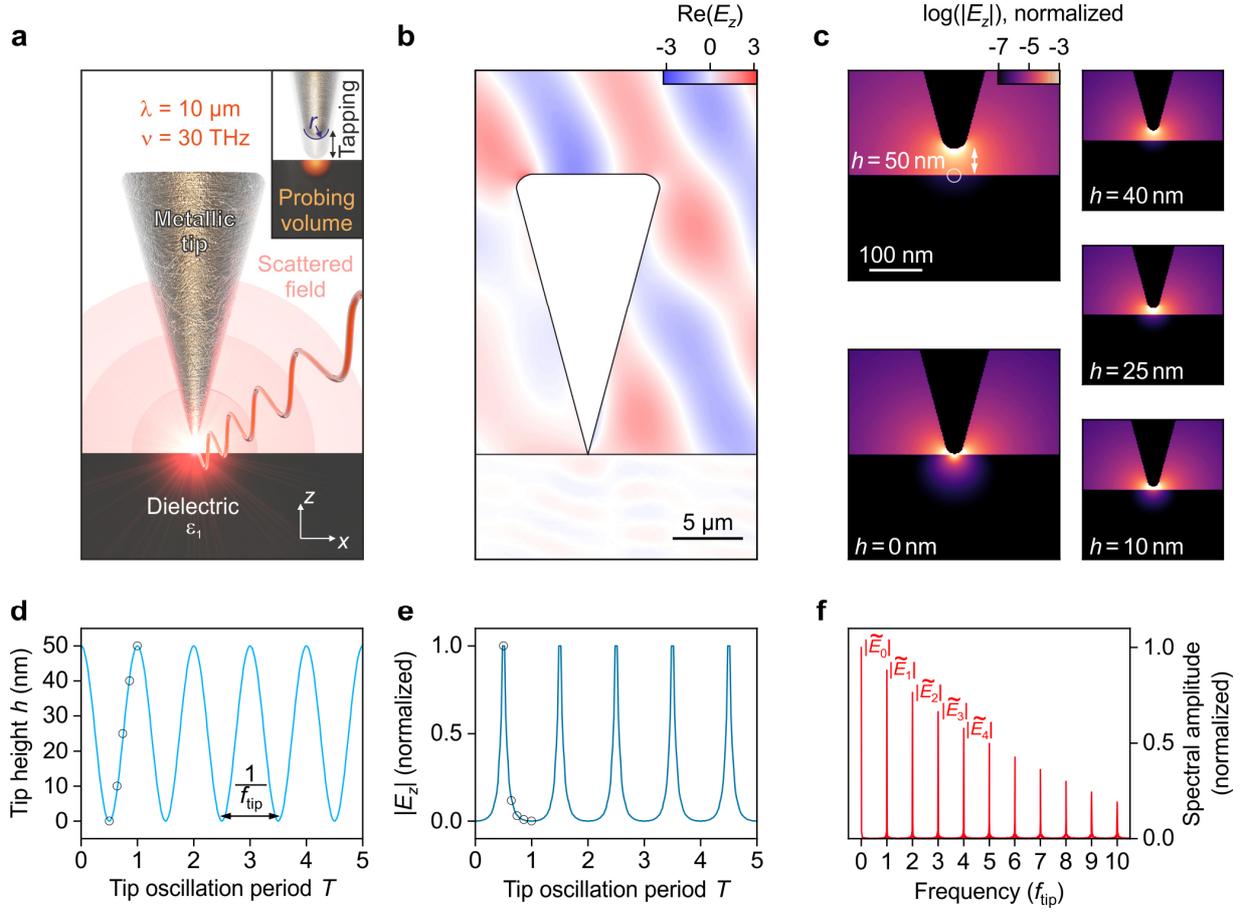

**Figure 1. Simulation geometry and Fourier demodulation analysis. a)** Sketch of the simulation layout including an oscillating metallic tip in close proximity to a sample with a dielectric function $\varepsilon_1$ (and a vanishing imaginary part of the dielectric function $\varepsilon_2 = 0$). The tip apex with radius $r$ is illuminated with mid-infrared radiation (frequency, $\nu = 30$ THz; wavelength, $\lambda = 10$ μm). The interaction between the tip in tapping motion and the sample takes place within the probing volume (see inset) and causes a scattering of the near fields within the nanojunction back into the far field. **b)** Real part of the out-of-plane electric field component $E_z$ as calculated by the finite element method. The black line indicates the surfaces of the tip and the sample. **c)** Close-up of the tip apex for various tip-sample distances $h$ highlighting the strong electric field enhancement within the junction. The data are normalized and plotted on a logarithmic scale. **d)** Sinusoidal modulation of the tip height *h* in time given in units of the tip oscillation period *T* corresponding to a frequency $f_{tip} = 1/T$. The black circles represent the tip heights *h* shown in **c**. **e)** Modulus of the out-of-plane electric field component $|E_z|$ at the sample surface beneath the tip apex (see white circle in **c**) calculated for the tip heights *h* in **d**. The black circles indicate the field strengths extracted for the distances shown in **c**. **f)** Fourier analysis of the electric field $E_z$ in **e** yielding the spectral components $\tilde{E}_n$ at the harmonic of order n of the tip oscillation frequency $f_{tip}$.

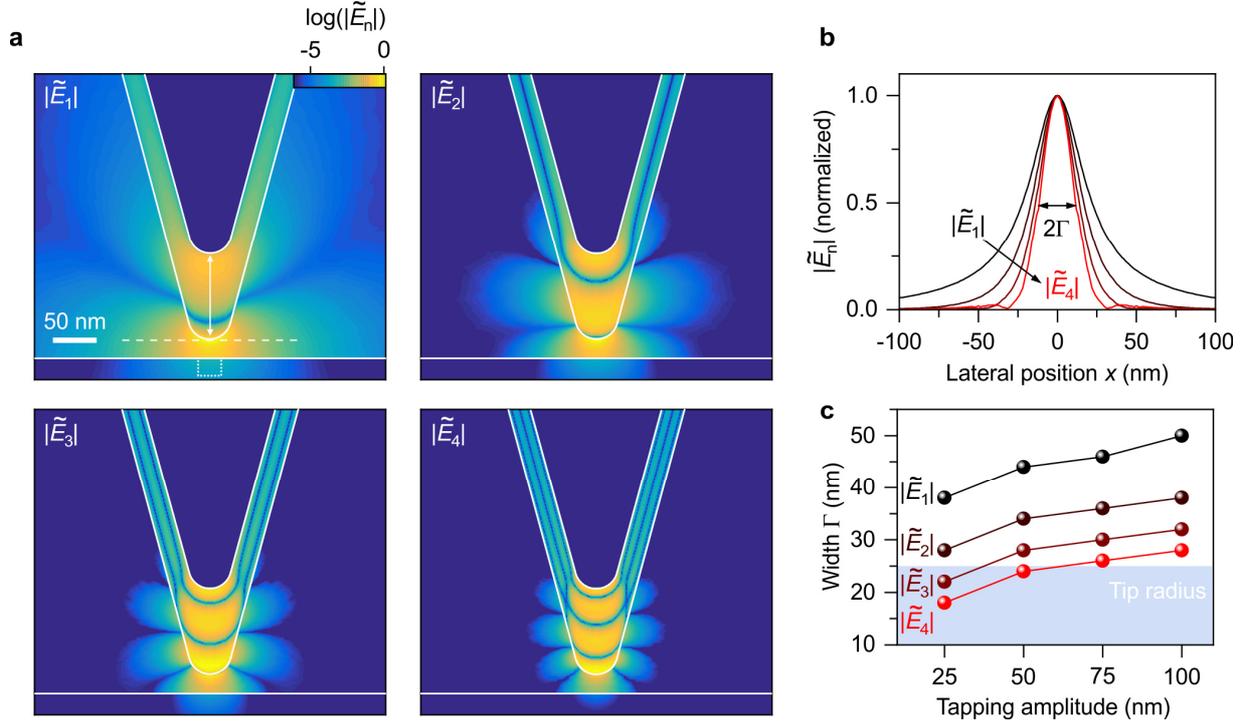

**Figure 2. Spatial distribution of the demodulated near fields. a)** Normalized maps of the field amplitude $|\tilde{E}_n|$ at demodulation orders n = 1 – 4 obtained by a Fourier analysis (compare Figure 1f) of the electric fields at each point in space with a tapping amplitude $A$ = 50 nm. The minimal tip-sample distance $d$ was set to 20 nm and the horizontal, white dashed line indicates where the line profiles in **b** were extracted. The solid white lines indicate the outline of the tip at the upper and lower points of inflection, and the sample surface, respectively. In the panel of $|\tilde{E}_1|$, the tapping motion of the tip is indicated by the white arrow and the highlighted region beneath the sample surface represents the area investigated in Figure 3. **b)** Line profiles with full width at half maximum $2\Gamma$ extracted 1 nm below the lowest point of the tip apex during a full oscillation cycle (see white dashed line in **a**) for different demodulation orders n. **c)** Widths $\Gamma$ of the field amplitudes $|\tilde{E}_n|$ below the apex (as extracted from **b**) for various tapping amplitudes. For a combination of high demodulation order and small tapping amplitude, the fields can even become localized on length scales smaller than the tip radius (see blue-shaded area).

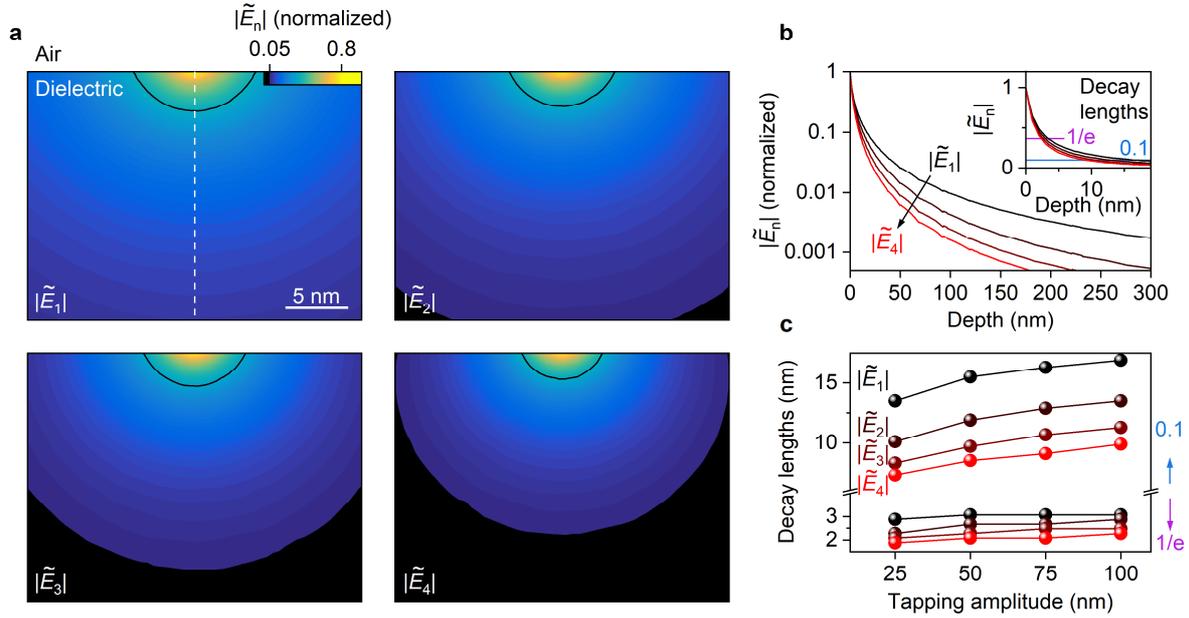

**Figure 3. Near-field probing volume within the sample. a)** Normalized field amplitudes $|\tilde{E}_n|$ for four different demodulation orders n extracted analogously to Figures 1f and 2a for a region close to the surface of the sample (see indicated region in Figure 2a). The data were obtained with a tapping amplitude $A = 50$ nm and a vanishing minimal tip-sample separation $d = 0$ nm. The solid black lines represent the 1/e contours of the decay of $|\tilde{E}_n|$. The white dashed line in the panel of $|\tilde{E}_1|$ indicates where the line cuts for panel **b** were extracted. **b)** Normalized line profiles extracted from the data in **a** along the indicated line. The inset shows the identical data on a linear scale and highlights the characteristic 1/e and 0.1 decay length. **c)** Decay lengths (at 1/e and 0.1 of the maximum value, see inset in Figure 3b) of the field amplitudes $|\tilde{E}_n|$ within the sample for a series of tapping amplitudes.

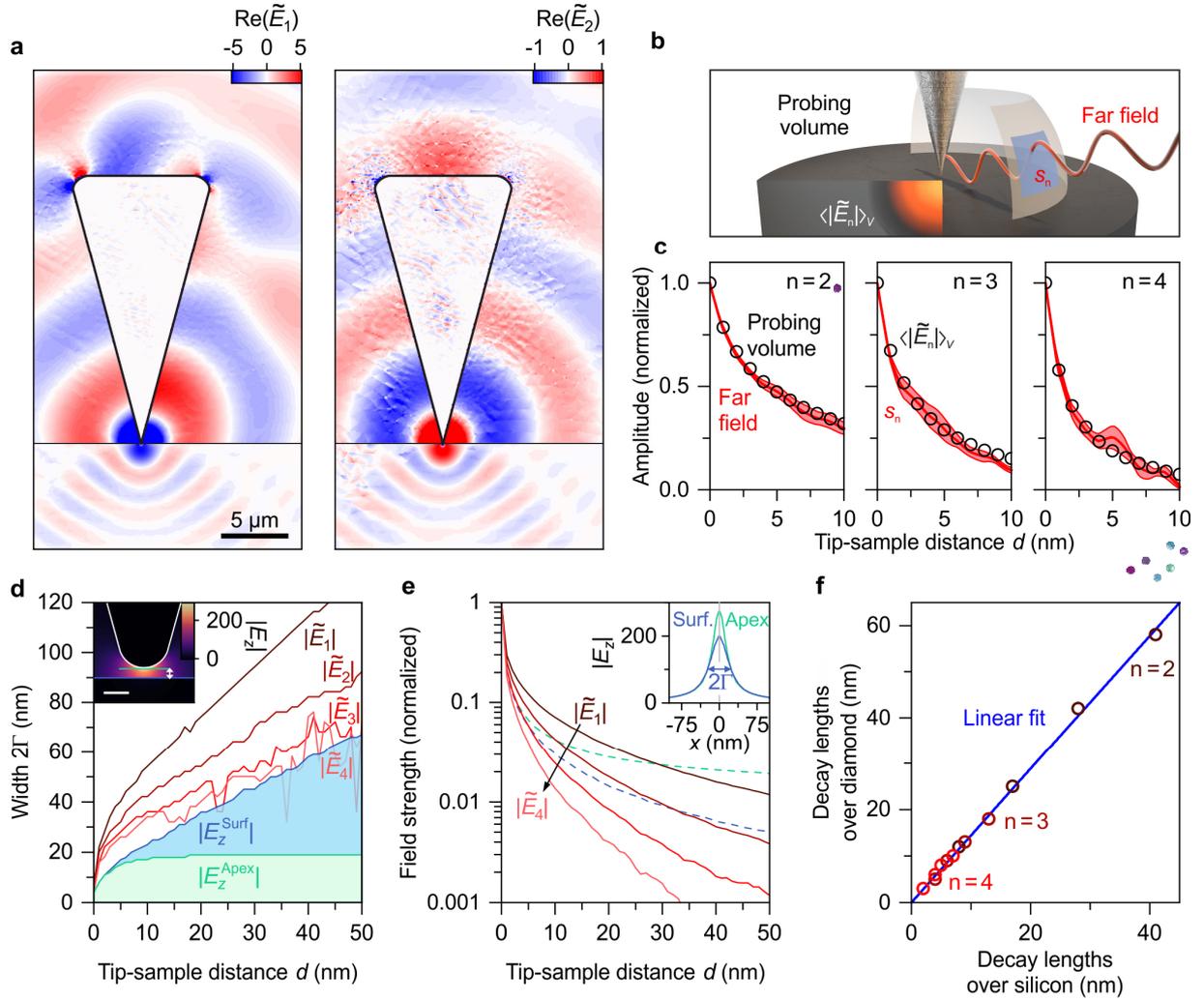

**Figure 4. Scattering to the far field at variable tip-sample distances. a)** Real parts of the electric fields Re($\tilde{E}_n$) at the first (left) and second (right) demodulation order obtained analogously to the images in Figure 2 with a vanishing minimal tip-sample separation $d = 0$ nm. Note that the values close to the tip apex strongly exceed the employed color scale by more than a factor of 100. **b)** Artist's view of the light scattering from the tip apex into the far field. The power of the scattered field passing through the surface of a dome is monitored as a function of the tip-sample distance to obtain the scattered amplitudes $s_n$, whereas the average near fields within the probing volume $V$ are given by $\langle|\tilde{E}_n|\rangle_V$. **c)** Mean field amplitudes $\langle|\tilde{E}_n|\rangle_V$ within the sample (circles) and the scattered far-field amplitudes $s_n$ (red-shaded regions) as a function of the tip-sample distance $d$. In the latter

case, the electric fields were evaluated on the surfaces of domes with different radii of up to 7 μm around the tip apex as visualized in **b**. The red-shaded regions indicate an interval of a standard deviation around the mean value of those data and include an alternative evaluation of the surface charge density of the tip (for details see Supporting Information). **d)** Width 2Γ of the distribution of the modulus of the out-of-plane electric field component $|E_z|$ (see inset in **d,e**) as a function of the tip-sample distance $d$ evaluated below the tip apex ($E_z^{\text{Apex}}$, green) and at the sample surface ($E_z^{\text{Surf}}$, blue). The width 2Γ of the demodulated field amplitudes $|\tilde{E}_n|$ (compare Figure 2b) is always determined at the sample surface. Inset: $|E_z|$ at a tip height $h = 10$ nm. The lines indicate where the field profiles at the sample surface (blue) and directly below the tip apex (green) were evaluated. White scale bar: 25 nm. **e)** Normalized field amplitudes $|\tilde{E}_n|$ evaluated at the sample surface directly underneath the tip apex for different tip-sample distances $d$. The green and blue dashed lines represent $|E_z|$ at the apex and the surface, respectively. Inset: Field profiles of $|E_z|$ extracted at the lines in the inset of **d** with $h = 10$ nm. The centers of the profiles are evaluated to obtain the data in the main panel of **e**. The tip is oscillating with $A = 25$ nm (**a,c**) and $A = 50$ nm (**d,e**), respectively. **f)** Comparison of the characteristic 1/e decay lengths of retraction curves over silicon (see **c**) and diamond (see Supporting Information) obtained for tapping amplitudes of 25 nm, 50 nm, 75 nm and 100 nm, and demodulation orders n = 2 – 4. The values obtained for identical probing parameters on the two materials exhibit a linear dependence as verified by the fit with a slope of 1.44.

**Methods**

All calculations in this work have been performed using a commercial finite element method software (COMSOL Multiphysics, including the radiofrequency (RF) package) to numerically solve Maxwell's equations for different tip-sample geometries. The simulation volume is $(28 \times 28 \times 42)$ μm³ in size and consists of two half-spaces, where the upper one ($z > 0$) is air ($\varepsilon_1^{Air} = 1$) and the lower one ($z < 0$) defines the sample. In this study, we investigated the dielectrics silicon ($\varepsilon_1^{Si} = 11.7$, ref. 61) and diamond ($\varepsilon_1^{dia} = 5.66$, ref. 62). The gold tip ($\varepsilon^{Au} = -3385 + 1457i$, ref. 63) was modelled by a cone in the upper half-space with a spherical apex. Rounded corners at the top suppress unrealistic field enhancement and minimize artefacts. If not stated otherwise, a length of 20 μm and a radius of curvature of $r = 25$ nm at the apex are used as the dimensions of the tip. Using the so-called "scattering problem" approach allowed us to distinguish between the illumination (background field) and the scattered field, which originates from the presence of the tip. To this end, we implemented the incident mid-infrared radiation (wavelength, $\lambda = 10$ μm; frequency, $\nu = 30$ THz) as a *p*-polarized plane-wave background field with an angle of incidence of 30° degrees with respect to the sample surface. We accounted for refraction and reflection using Snell's law and Fresnel's equations. Perfectly matched layers surrounding the simulation volume on all sides dampen the outgoing radiation and thereby prevent any reflections from the simulation boundaries. To accurately simulate the near field in the tip-sample junction, the mesh in and close to this region is chosen extremely fine (with element sizes down to 0.5 nm). Additionally, the mesh nodes at the air-sample-interface have been fixed for all tip heights *h* to ensure a reliable demodulation of the fields in this critical area. The numerical analysis and demodulation has been performed as a post-processing step using MATLAB.

## ASSOCIATED CONTENT:

**Supporting Information**

The Supporting Information is available free of charge on the ACS Publications website at DOI: Simulation details and far-field quantities, electric field decay lengths for different tip radii and materials, deformation of the probing volume for finite tip-sample distance, comparison of retraction curves on different materials, comparison with the finite-dipole model, and demodulated fields at the surface (PDF)

## AUTHOR INFORMATION:

**Corresponding Author**

*E-Mail: fabian.mooshammer@ur.de, Fax: +49 941 943 4223

*E-Mail: rupert.huber@ur.de, Fax: +49 941 943 4223

**Author Contributions**

F.M. and M.A.H. contributed equally to this work (‡). F.M., M.A.H. and R.H. conceived the study. F.M., and M.A.H. performed the simulations and evaluated the data with input by F.S., M.P., M.Z., and R.H.. F.M. and F.S. performed the experiments. F.M., M.A.H and R.H wrote the manuscript with input from all authors. All authors contributed to the discussions and have given approval to the final version of the manuscript.

**Notes**

The authors declare no competing financial interest.

ACKNOWLEDGEMENTS:

The authors thank Alexander Neef and Tyler L. Cocker for fruitful discussions. The work was supported by the European Research Council through Grant Number 305003 (QUANTUMsubCYCLE) as well as by the Deutsche Forschungsgemeinschaft (through grant numbers HU1598/3 & CO1492, SFB 1277 Project A05, and GRK 1570).